\documentclass[showpacs,epsfig,preprintnumbers,amsmath,amssymb,nofootinbib,20pt]{revtex4-1}
\pdfoutput=1
\usepackage{graphicx}

\usepackage{graphicx}
\usepackage{dcolumn}
\usepackage{bm}

\begin{document}

\def\d{{\rm d}}
\def\Epos{E_{\rm pos}}
\def\ap{\approx}
\def\eff{{\rm eft}}
\def\L{{\cal L}}
\newcommand{\vev}[1]{\langle {#1}\rangle}
\newcommand{\CL}   {C.L.}
\newcommand{\dof}  {d.o.f.}
\newcommand{\eVq}  {\text{EA}^2}
\newcommand{\Sol}  {\textsc{sol}}
\newcommand{\SlKm} {\textsc{sol+kam}}
\newcommand{\Atm}  {\textsc{atm}}
\newcommand{\Chooz}{\textsc{chooz}}
\newcommand{\Dms}  {\Delta m^2_\Sol}
\newcommand{\Dma}  {\Delta m^2_\Atm}
\newcommand{\Dcq}  {\Delta\chi^2}
\newcommand{\nbb}{$\beta\beta_{0\nu}$ }
\newcommand {\be}{\begin{equation}}
\newcommand {\ee}{\end{equation}}
\newcommand {\ba}{\begin{eqnarray}}
\newcommand {\ea}{\end{eqnarray}}
\def\VEV#1{\left\langle #1\right\rangle}
\let\vev\VEV
\def\e6{E(6)}
\def\10{SO(10)}
\def\21{SA(2) $\otimes$ U(1) }
\def\321{$\mathrm{SU(3) \otimes SU(2) \otimes U(1)}$ }
\def\lr{SA(2)$_L \otimes$ SA(2)$_R \otimes$ U(1)}
\def\422{SA(4) $\otimes$ SA(2) $\otimes$ SA(2)}
\newcommand{\AHEP}{%
School of physics, Institute for Research in Fundamental Sciences
(IPM)\\P.O.Box 19395-5531, Tehran, Iran\\

  }
\newcommand{\Tehran}{%
School of physics, Institute for Research in Fundamental Sciences (IPM)
\\
P.O.Box 19395-5531, Tehran, Iran}
\def\roughly#1{\mathrel{\raise.3ex\hbox{$#1$\kern-.75em
      \lower1ex\hbox{$\sim$}}}} \def\lsim{\roughly<}
\def\gsim{\roughly>}
\def\ltap{\raisebox{-.4ex}{\rlap{$\sim$}} \raisebox{.4ex}{$<$}}
\def\gtap{\raisebox{-.4ex}{\rlap{$\sim$}} \raisebox{.4ex}{$>$}}
\def\lsim{\raise0.3ex\hbox{$\;<$\kern-0.75em\raise-1.1ex\hbox{$\sim\;$}}}
\def\gsim{\raise0.3ex\hbox{$\;>$\kern-0.75em\raise-1.1ex\hbox{$\sim\;$}}}

\title{Pico-charged intermediate particles rescue dark matter interpretation of 511 keV signal}
\date{\today}

\author{Y. Farzan}\email{yasaman@theory.ipm.ac.ir}
\author{M. Rajaee}\email{meshkat.rajaee@ipm.ir}
\affiliation{\Tehran}
\begin{abstract}
Various alleged  indirect dark matter search signals, such as the 511 keV line from galaxy center or the PAMELA/AMS02 signal, are often challenged by the absence of corresponding signal from dwarf galaxies and/or by the absence of an impact on CMB through delayed recombination. We propose a novel scenario that can avoid these bounds based on the decay of dark matter, $X$, to a pair of intermediate particles $C$ and $\bar{C}$ with a lifetime much greater than the age of universe. The annihilation of these intermediate particles eventually leads to a dark matter signal. The bounds from CMB can be easily avoided by the fact that at the time of recombination, not enough $C$ particles had been accumulated.  In order to keep $C$ particles from leaving the galaxy, we assume the particles have a small electric charge so in the galactic disk, the magnetic field keeps the $C$ particles in the vicinity of their production. However, they can escape the dwarf galaxies and the dark matter halo where the magnetic field is weak, leading to null signal from these regions. The small charge can have interesting consequences including a signal in direct dark matter search experiments.
\end{abstract}

\date{\today}
\maketitle
\section{Introduction}
From cosmological and astronomical observations, the evidence for the existence of a new form of matter known as Dark Matter (DM) is overwhelming. To establish the existence of DM and to learn its microscopic properties, a host of direct and indirect DM search experiments have been designed. A number of these experiments have found signals that might originate from DM but none of them has been so far conclusively established as DM discovery. Some of these alleged signals have disappeared altogether by collecting more data. However, there are other examples that have stood the test of time and statistics but the dark matter interpretation of these signals has become less popular only
because the simplest DM scenario accounting for the observed signal also predicts other accompanying signals that have not been observed.
The notorious examples of such accompanying signals are delayed recombination (and therefore impact on CMB) and  a signal  from dwarf satellite galaxy where the ratio of DM to ordinary matter is expected to be high. The aim of the present paper is to propose a scenario that can avoid these bounds, rendering the DM interpretation of observed signals viable.

For over 40 years, a signal of 511 keV photon line has been observed from galaxy center \cite{Siegert:2015knp}. The 11 years data from the SPI  spectrometer at INTEGRAL established the signal at 56 $\sigma$ C.L. The 511 keV photons are expected to come from annihilation of the electron positron pair in form of positronium atoms in the bulge of galaxy. The origin of the positrons is a mystery. A plausible source can be the annihilation of DM.  Although other more conventional sources such as $X$-ray binaries and pulsars can also contribute to positrons but they should also lead to higher energy photons with a characteristic  spectral shape that is not compatible with observations \cite{Bandyopadhyay:2008ts,Muno:2004bs}. Moreover, these point sources predict a  morphology  different from  that of the observed  $511$ keV line with bulge to disk ratio larger than one \cite{ast}.

As has been shown in \cite{Boehm:2003bt}, if DM is composed of particles with mass of few MeV annihilating to $e^-e^+$ with a cross-section of $\sim 10^{-4}$~pb, the intensity and morphology of the 511 keV line can be neatly explained.  As indicated in \cite{Wilkinson:2016gsy}, if the annihilation to $e^-e^+$ is
a $S$-wave  process, the reioinzing energy dump at recombination will be large enough to lead to delayed recombination so the scenario is ruled out by CMB . On the other hand, if the annihilation is $P$-wave, at temperatures of order of DM mass, we expect an annihilation cross section of  $\sim$10 pb which washes out the relic density of DM. Thus, this simple scenario is ruled out.

We suggest a scenario that can circumvent the bound from CMB. Let us suppose that DM is composed of scalar particles of mass $\sim 10$ MeV
denoted by $X$.
Similarly to the SLIM scenario, $X$ particles can be produced via thermal freeze-out scenario with $\sigma (XX\to \nu \bar{\nu})\sim 1$~pb \cite{Boehm:2006mi,Farzan:2009ji}.
Let us suppose that the $X$ particles decay with a lifetime larger than the age of universe to a pair of $C$ and $\bar{C}$ particles which have an electric charge of $|q|\stackrel{>}{\sim} 10^{-12} $. The magnetic field in galaxy will be enough to keep the $C$ particles at a distance  from galaxy center close to that at their production point. Thus, the profiles of $C$ and $\bar{C}$ particles will follow the profile of the dark matter in galaxy: $n_C=n_{\bar{C}}= f n_X$ where
$f=\Gamma_X t^0$ in which $\Gamma_X$ is the decay rate of $X$ and $t^0$ is the age of the Universe. If  the $C\bar{C}$ annihilation leads to the $e^+$ production, the morphology of the 511 keV line will be similar to DM annihilation scenario (with $n_X^2$ dependence) which, as mentioned before, fits the observation. Notice that the $C$ and $\bar{C}$ particles will be relativistic so $f$ cannot be larger than 1 \% \cite{Audren:2014bca}. At the recombination, the $C$ and $\bar{C}$ density will be too small  to lead to any significant reionization so the bound from CMB is circumvented. Moreover, because of the smallness of the magnetic field, the dwarf galaxies  are not in general
expected to accumulate the $C\bar{C}$ pairs so we do not in general expect a significant 511 keV signal from a typical dwarf galaxy with small magnetic field \cite{Siegert:2016ijv}. As shown in \cite{Siegert:2016ijv}, among the $39$ dwarf galaxies of milky way that were studied,  only  Reticulum II  shows a $511$ keV signal with  a significance larger than $3\sigma$. Unfortunately the magnetic field in this dwarf galaxy is not properly known. Better
determination of the magnetic field in dwarf galaxies will provide a new tool to test our model.

In section II, the scenario is explained and various bounds on its parameters is discussed. In section III, the prospect of testing the scenario by direct dark matter search experiments is discussed. In section IV, the results are summarized.

\section{The scenario}
We assume that dark matter is mainly composed of particles, $X$. The metastable $X$ particles decay into $C\bar{C}$ with a decay rate of $\Gamma_X$. If these $C$ particles remain at a distance  from galaxy center close to that at their production point, their number density now (at $t^0$) will be given by
\be n_C=\frac{\rho_X}{m_X} f \ \  \ \ {\rm where} \ \ \ \ f=\Gamma_X t^0.\ee $f$ is the fraction of $X$ particles that have decayed.
Observed intensity of $511$ keV line from galactic center after eleven years of data taking by INTEGRAL is  $  \Phi_{511} = (0.96 \pm 0.07)\times 10^{-3} \rm{ph} ~\rm{cm^{-2}}~\rm{sec^{-1}} $ \cite{Siegert:2015knp} which is in remarkably good agreement with earlier reports \cite{Boehm:2003bt}. Let us calculate expected flux from galactic center in our model.
Assuming that each $C\bar{C}$ pair annihilation leads to $N_{pair}$ pairs of $e^-e^+$ and therefore to $N_{pair}$ pairs of photons,
the flux from the galaxy bulge can be estimated as
\begin{equation}
\Phi _{511} \simeq 2  N_{pair} \frac{\int _0 ^{r_b} (n_X(r)) ^2 f^2  \langle \sigma _{C \bar{C}} v \rangle 4 \pi r^2 dr}{4 \pi r_{sol}^2}
\end{equation}
where $r_b\simeq 1$ kpc is the bulge radius, $r_{sol} \simeq 8$ kpc is the distance of the solar system to the galaxy center and finally $n_X ={\rho_X (r)}/{m_X}$ is dark matter number density. Considering an NFW halo profile  with a scale radius of 20 kpc and a local dark matter density of  0.4 ${\rm GeV/ cm^3}$ \cite{Hooper:2016ggc}, we find that in order to reproduce the observed intensity of the 511 keV line \cite{Boehm:2003bt}, the $C\bar{C}$ annihilation cross section should satisfy
\be \sigma_{C\bar{C}}=50~ pb\left(\frac{1.5\times 10^{-4}}{f}\right)^2 \left(\frac{m_X}{5~{\rm MeV}}\right)^2\frac{1}{N_{pair}} .\ee


 If $C\bar{C}$ directly annihilate to $e^-e^+$, such
large annihilation cross section requires a value of coupling to the electron that has already been ruled out. (For example with effective coupling $(\bar{e}e)(\bar{C} C)/\Lambda$,  the cross section of $50$ pb requires $\Lambda \sim 100$ GeV which means  the virtual particle  leading to this effective coupling should have already been discovered at LEP.) The $C\bar{C}$ pair can annihilate to a pair of new scalars $\phi \bar{\phi}$ which in turn decay into $e^-e^+$ ($\phi \to e^-e^+$). Thus, in this model $N_{pair}=2$.
 An effective coupling  of form \be g_\phi \phi \bar{e}e\ee  can  lead to $\phi \to e^-e^+$. For $g_\phi <10^{-11}$, the production of $\phi$ in the supernova core will be too small to affect its cooling so the supernova bounds  can be satisfied. Moreover, taking $g_\phi>10^{-15}$, $\phi$ can decay before traveling a distance exceeding 10~pc.  If $C$ is confined within a distance smaller than $\sim 100$ pc from its production point and if $\phi$ travels less than $\sim$100 pc before decay, the morphology of 511 keV line will remain consistent with a $n_{DM}^2$ profile which is favored by observation.\footnote{The angular resolution of the INTEGRAL observatory is $2^{\circ}$ which corresponds to a distance of $  (2 \pi/180)\times (8 ~kpc)= 280~ pc$ in the galactic center.} We therefore take $0.3 \times 10^{-15}< g_\phi < 10^{-11}$.  Since we do not want $\phi$ to have electroweak interaction, we take it to be a singlet of the Standard Model (SM) gauge group. The $g_\phi$ coupling can originate from mixing with the SM Higgs via the term $$ a_\phi \phi |H|^2$$ which leads to \be  g_\phi = \sqrt{2}\frac{a_\phi v}{m_h^2 }Y_e\ee in which $Y_e$ is the Yukawa coupling of the SM Higgs to the electron. Since $a_\phi\stackrel{<}{\sim}m_\phi$, at tree level no fine tuning is required to maintain the required hierarchy between the $H$ and $\phi$ masses. In this sense, the model is natural.  The number density of $n_\phi$ produced via $e^-e^+ \to \phi \gamma$ in the early universe can be estimated as $n_\phi \sim \langle \sigma_\phi v \rangle n_e^2 H^{-1} |_{T=m_\phi}$ where $\sigma_\phi=e^2 g_\phi ^2/8 \pi m_\phi^2$. Thus, for the time range after $T=m_\phi$ and before $\Gamma_\phi^{-1}$, we can write $n_\phi/n_\gamma=8 \times 10^{-8} (g_\phi/10^{-11})^2$.  Of course the density of $\phi$ at $T\sim 1$ MeV was too small to affect BBN. Moreover for $g_\phi >10^{-15}$, $\phi$ particles decay well before recombination. {At the time of decay [at $t=\Gamma_\phi^{-1}$ corresponding to $T=(m_\phi M_{Pl}^*/(4\pi))^{1/2}g_\phi$], we have $\rho_\phi/\rho_\gamma\sim (n_\phi/n_\gamma)(m_\phi/T)\sim 5 \times 10^{-7} (g_\phi/10^{-11})\ll 1$ so the decay of $\phi$ cannot pump enough energy to warm up the background plasma and affect CMB. }


Let us now build a mechanism to keep $C$ particles confined.  There are (at least) two possibilities:
\begin{itemize}
\item The $C$ particles have velocities smaller than the escape velocity; {\it i.e.,} $v\stackrel{<}{\sim} 10^{-3}$. In order for $C$ particles to be so slow at production, the fine tuned relation $(m_X-2m_C)/m_C \sim v^2<10^{-6}$ should be satisfied which is difficult to justify. Another possibility is that the $C$ particles somehow lose  enough energy before traveling a distance of $\sim$ 10 pc.
\item  The $C$ particles  stay in galaxy due to the interaction of their tiny electric charge, $q$, with the background magnetic field. In the galactic center, there is a magnetic field of axial symmetry with a magnitude of $\sim$10~$\mu$G \cite{LaRosa:2005ai,Beck:2008ty,Chuzhoy:2008zy}
 which  keeps the $C$ particles confined   in the galactic center. To reproduce the observed morphology of the 511 keV line, the Larmour radius $r_L=5~pc \times (\frac{3\times 10^{-11}}{q}) (\frac{ m_X}{5 ~\rm{MeV}} ) (\frac{10 ~\mu \rm{Gauss}}{B})$ in the galaxy center magnetic field should be smaller than  100~pc. The required value of $q$ versus $m_X=2 E_C$ is shown in Fig 1-a.
With such small $q$, all the relevant bounds are satisfied but for the parameter range of our interest, as shown in \cite{Chuzhoy:2008zy}, the pico-charged particles can become accelerated in the supernova shock waves within time scales of 100 Myr so unless an energy loss mechanism with $\tau_E^{-1}=d \log E/dt$ greater than $(100 ~{\rm Myr})^{-1}$ is at work, the Larmours radius  will increase  and $C$ and $\bar{C}$ particles will escape the galactic center.  The synchrotron  radiation  from $C$ particles in magnetic field is suppressed by $q^4$ and is too small to efficiently cool down the $C$ particles.
\end{itemize}
In summary, in both cases we need a mechanism for energy loss. In what follows we will focus on the second solution with pico-charged $C$ trapped in the galactic magnetic field. Fortunately, as we shall see below, the same mechanism that provides electric charge for $C$ also naturally provides a mechanism for cooling.

Let us now briefly discuss how the $C$ particles can acquire such small (but non-zero) electric charge. As shown in \cite{Feldman:2007wj}, the tiny electric charge can come from the mixing of hyper-charge gauge boson with the gauge boson of a new $U(1)_X$ gauge symmetry under which the $C$ particles are charged. After breaking of $U(1)_X\times SU(2)\times U(1)_Y\to U(1)_{em}$, the $W$ and $Z$ bosons as well as the new gauge boson become massive. Moreover, the particles charged under $U(1)_X$ obtain a tiny charge under $U(1)_{em}$, given by the mixing between $U(1)_X$ and $U(1)_Y$ gauge bosons.
The new gauge boson is often called dark photon and denoted by $\gamma^\prime$. The new gauge coupling is denoted by $g_X$; {\it i.e.,} The coupling of $C$ and $\bar{C}$ to $\gamma^\prime$ is $g_X$.
Interestingly, as shown in appendix A of \cite{Feldman:2007wj}, if a certain relation between kinetic and mass mixing holds, independently of the mass of new gauge boson and its coupling $g_X$, relation $q=g\epsilon/e$ holds and the mass of $\gamma^\prime$ becomes arbitrary. Moreover,
  at the tree level, $\gamma^\prime$  does not couple to the SM fermions.

  Notice that there is no symmetry to prevent $\gamma^\prime$ particles from decay.
  Kinematically available decay modes are $\gamma \gamma, \ \gamma \gamma \gamma, \ \nu \bar{\nu}$ and etc. Notice that according to the Landau-Yang theorem, $\gamma^\prime \to \gamma \gamma$ is forbidden \cite{Yang:1950rg}.
For general values of kinetic and mass mixing, the coupling of the electron and other SM charged fermions to $\gamma^\prime$ will be of order of $q$, too. An electron loop then leads to $\gamma^\prime \to \gamma \gamma \gamma$ with a lifetime of $\sim 100000$ years. However, as mentioned above there is a limit where SM charged particles  do not couple to $\gamma^\prime$. In this limit, the dominant diagram leading to decay $\gamma^\prime \to \gamma \gamma \gamma$ will be a one-loop diagram  in which $C$ particles propagate. The lifetime will be then given by $\sim (4\pi (16\pi^2)^2/(m_{\gamma^\prime}g_X^2(q^2)^3))\sim 7 \times 10^{45}~{\rm years} \gg t^0$. Another decay mode is $\gamma^\prime \to \gamma Z^* \to \gamma \nu \bar{\nu}$ which will have even smaller rate $\sim g_X ^2 m_{\gamma^\prime}q^4(m_{\gamma^\prime}^2/m_{Z^\prime}^2)^2/(4\pi(16\pi^2)^2)$. For simplicity, we will assume that the relation between kinetic and mass mixing holds, leading to a stable $\gamma^\prime$ which, as we see below, will also provide a cooling mechanism.

For $T\gg m_C$, the cross-section of ${\rm SM}+{\rm SM} \to C+\bar{C}$ (via $s$-channel photon exchange) is $\sigma\sim q^2e^2/(4\pi T^2)$ which means at temperature $T$, the ratio of the number density of electromagnetically produced $C$ (via processes such as $e^+e^- \to C\bar{C}$) to $n_\gamma$ is $\dot{n}_CH^{-1}/n_\gamma=5 \times 10^{-5}(q/10^{-11})^2$ which is still too large as it can lead to sizeable reionization at  recombination. Taking $\gamma^\prime$ lighter than $C$ particles, the  $C\bar{C} \to \gamma^\prime \gamma^\prime$ annihilation can efficiently convert the $C\bar{C}$ pairs to $\gamma^\prime \gamma^\prime$. In fact, solving the Boltzman Equation, $\dot{n}_C+3 Hn_C=-\langle \sigma( C\bar{C} \to \gamma^\prime \gamma^\prime) v \rangle n_C^2$,  we find for $\langle \sigma( C\bar{C} \to \gamma^\prime \gamma^\prime) v \rangle n_CH^{-1} |_{T=m_C}\gg 1$ (which can be satisfied if $\langle \sigma( C\bar{C} \to \gamma^\prime \gamma^\prime) v \rangle \gg 10^{-39} {\rm cm}^2$),
\be \label{IniC} n_C=\frac{T^3}{M_{Pl}^* \langle \sigma( C\bar{C} \to \gamma^\prime \gamma^\prime) v \rangle m_C} \ \  \ {\rm at} \ \ \  T\ll m_C. \ee
 Ref. \cite{Slatyer:2016qyl}, has derived a bound of $\sim 10^{-30}~\rm{cm^{3} sec^{-1}}$ on $\langle \sigma( DM+DM \rightarrow e^+ e^-) v \rangle$ from reionization effect on CMB for dark matter mass in the range of $0.5$ MeV to $15$ MeV. In  models that dark matter pair directly annihilate into $e^+ e^-$, the energy dump rate per volume is proportional to $n_{DM}^2 \langle \sigma (DM+DM \rightarrow e^+ e^-) v \rangle$. In our case, the energy dump rate per volume from $C \bar{C}$ annihilation is proportional to $n_C^2 \langle \sigma (C \bar{C} \rightarrow \phi \phi) v \rangle$. As a result if $(\frac{n_C}{n_{DM}})^2  \langle \sigma (C \bar{C} \rightarrow \phi \phi) v \rangle < 10^{-30} ~\rm{cm^{3} sec^{-1}}$, the bounds from CMB will be satisfied. This condition yields
 \be
 {\langle \sigma (C \bar{C} \rightarrow \gamma^\prime \gamma^\prime)  v \rangle} < 2 \times 10^{-7}  {\rm b}     \sqrt{\frac{\langle \sigma (C \bar{C} \rightarrow \phi \phi) v  \rangle  } {100   ~{\rm pb}}}.
 \ee
As we shall see this bound can be readily satisfied for the parameter range of our interest.
From Eq. (\ref{IniC}) we observe that  for large enough $\langle \sigma( C\bar{C} \to \gamma^\prime \gamma^\prime) v \rangle$, all $C \bar{C}$ pairs in the early universe will practically convert to $\gamma^\prime$ pairs well before recombination: \be \frac{n_{\gamma^\prime}}{n_\gamma}=10^{-4}\left(\frac{q}{10^{-11}}\right)^2.\ee
The $\gamma^{\prime}$ particles will be non-relativistic at matter-radiation equality (when structure formation practically starts)
and at recombination so they count as components of cold dark matter. As a result the bound on relativistic degrees of freedom from CMB does not apply for $\gamma^\prime$; However,  As long as $m_{\gamma^\prime}<$~MeV, these particles at BBN act as relativistic degrees of freedom so BBN bound $\Delta N_{eff}<0.7$ \cite{Kamada:2015era} should be imposed. To be on safe side, we take $n_{\gamma^\prime}/n_\gamma<0.1$ which implies $q<3 \times 10^{-10}$. Remembering $\rho^0_{DM}={\rm keV}{\rm cm}^{-3}$ and $n_\gamma=400~{\rm cm}^{-3}$, as long as $m_{\gamma^\prime}<10~{\rm keV} (10^{-11}/q)^2$, the contribution of $\gamma^\prime$ to DM will be subdominant. Thus, for $10^{-11}<q<3\times 10^{-10}$, we arrive at  $10~{\rm eV}<m_{\gamma^\prime}<10~{\rm keV}$. For massive $\gamma^\prime$, we expect the density of $\gamma^\prime$ in our galaxy center  to be larger than the average $\gamma^\prime$ density of the universe: \be n_{\gamma^\prime}|_{GC} =\frac{\rho_{DM}|_{GC}}{\langle \rho_{DM}\rangle} \langle n_{\gamma^\prime}\rangle.\ee Dark matter density  in the galactic bulge is still uncertain. Even if we fix the local dark matter density to 0.4 GeV/cm$^3$ and assume generalized NFW model  with arbitrary  $\gamma$ parameter \cite{Hooper:2016ggc}, we find that at the outer layers  of the galactic bulge ({\it i.e.,} at $r=1~{\rm kpc}$), $\rho_{DM}\sim 3-10 ~{\rm GeV}~{\rm cm}^{-3}$. Assuming a nominal and rather conservative value $\rho_{DM}|_{GC}  = 3 ~{\rm GeV}~{\rm cm}^{-3}$ at bulge, we find that $n_{\gamma^\prime}$ at Galaxy Center (GC) is $n_{\gamma^\prime}\sim 10^5 (q/10^{-11})^2{\rm cm}^{-3}$.  Increasing dark matter density  value to $\rho_{DM}|_{GC}  = 10 ~{\rm GeV}~{\rm cm}^{-3}$, the cooling process for $C$ particles becomes even more efficient. On average, the relativistic $C$ particles at each collision with background $\gamma^\prime$ can lose  energy of $$\Delta E_C=m_{\gamma^\prime} \left(\frac{E_C}{m_C}\right)^2 v^2.$$  The scattering cross section of $C$ off $\gamma^\prime$ is $\sigma_S\simeq 3g^{ 4}_X /(4 \pi m_C^2)$ so the cooling time, $\tau_E$, down to final velocity $v_f$ is given by
\be \tau_E= \int_{m_C(1+v_f^2/2)}^{m_X/2}\frac{dE_C}{\Delta E_C} \frac{1}{\sigma_S v n_{\gamma^\prime}}|_{GC}\simeq  100~{\rm Myr} \left(\frac{m_C}{5 ~{\rm MeV}}\right)^3 \left(\frac{10~{\rm keV}}{m_{\gamma^\prime}}\right) \left( \frac{0.03}{v_f} \right)\left(\frac{0.15}{g_X}\right)^4\left(\frac{10^5~{\rm cm}^{-3}}{n_{\gamma^\prime}}\right). \ee
It is  remarkable that $\tau_E$ is given by $(n_{\gamma^\prime} m_{\gamma^\prime} )^{-1}$ so irrespective of $q$, as long as $n_{\gamma^\prime} m_{\gamma^\prime}$ is fixed, we obtain the same $\tau_E$.
Notice that for $v\sim 1$, the energy loss rate is high but as $v$ decreases, the energy loss rate decrease as $v^3$.
 When $v_f$ reaches
 \be \label{vf} v_f=0.3 \left( \frac{0.15}{g_X}\right)^4 \left(\frac{10^4 {\rm cm}^{-3}}{n_{\gamma^\prime}} \right) \left( \frac{m_C}{5 ~{\rm MeV}}\right)^3 \left(\frac{m_{\gamma^\prime}}{10~{\rm keV}}\right),\ee
 the energy gain from supernova shocks compensates the energy loss from scattering. In the vicinity of the Sun, $n_{\gamma^\prime}$, which follows the $\rho_{DM}$ profile, is 10 times smaller than $n_{\gamma^\prime}$ at the galaxy center.  That is $n_{\gamma^\prime}|_{\rm at ~ Earth}=10^4 (q/10^{-11})^2 {\rm cm}^{-3}$. From Eq. (\ref{vf}), we therefore  expect the velocity of $C$ particles around us to be $\sim {\rm few} ~0.1$.

Notice that the $\gamma^\prime$ particles after scattering gain a recoil energy which can cause them to escape the galaxy. That means over time, the density of $\gamma^\prime$ in galaxy decreases. Let us check if during the lifetime of the galaxy ({\it i.e.,} $\sim t^0\sim 10$~Gyr), enough $\gamma^\prime$ remains to cool down all the produced $C$ and $\bar{C}$. We have chosen the value of $g_X$ such that during 0.1 Gyr, the number of scatterings of each $C$ and $\bar{C}$ to be $$\delta n_{\gamma^\prime/C}\sim \int_{m_C(1+v_f^2/2)}^{m_X/2} \frac{dE_C}{ \Delta E_C}=\frac{m_C}{2 m_{\gamma^\prime}} \log \left[ \frac{m_X/2-m_C}{m_Cv_f^2/2} \frac{2m_C}{m_X/2+m_C}\right] .$$
 Thus, during $t^0=10$ Gyr, each $C$ and $\bar{C}$ repels $2 (t^0/0.1~{\rm Gyr})\delta n_{\gamma^\prime/C}\sim 10^5 ({\rm keV}/m_{\gamma^\prime})$ dark photon from the galaxy center. As a result, the number of $\gamma^\prime$ repelled from unit volume is $\Delta n_{\gamma^\prime}=200000 f(\rho_X|_{GC}/m_X)({\rm keV}/m_{\gamma^\prime})$. Requiring $\Delta n_{\gamma^\prime}< n_{\gamma^\prime}|_{GC}$ implies \be f<10^{-2}\left( \frac{q}{10^{-11}}\right)^2 \left(\frac{m_{\gamma^\prime}}{10~{\rm keV}}\right) \left(\frac{m_{X}}{10~{\rm MeV}}\right).\label{f}\ee
  It is remarkable that the dependence of $\delta n_{\gamma^\prime/C}$ on $m_X$ is mild so even with $m_X\gg m_C$, the number of repelled $\gamma^\prime$ by  each $C$ and $\bar{C}$ does not significantly increase and we arrive at Eq. (\ref{f}).
  Remember that $\Omega_{\gamma^\prime}$ sets an upper bound on $m_{\gamma^\prime} q^2$. The bound in Eq. (\ref{f}) on $f$ is comparable to the bound from the constraint on the relativistic component of DM \cite{Audren:2014bca}.
Another upper bound on $f$ comes from the bound on the rate of $C\bar{C}\to \gamma^\prime \gamma^\prime$ in the galaxy. We want the rate of this process to be much smaller than $t_0^{-1}$ so that $C$ and $\bar{C}$ pairs can be accumulated. This yields $n_C \langle \sigma(C\bar{C} \to \gamma^\prime\gamma^\prime) v \rangle t_0 \ll 1$. Since $\langle \sigma (C\bar{C}\to \gamma^\prime \gamma^\prime) v \rangle \sim 2\times 10^{-3} ~{\rm b}(g_X/0.2)^4 (5~{\rm MeV}/m_C)^2 $, this bound again leads to \be \label{waisted} f <6\times 10^{-4}\left( \frac{m_X}{10~{\rm MeV}}\right)\left(\frac{0.15}{g_X}\right)^4 \left( \frac{m_C}{5~{\rm MeV}} \right)^2. \ee

Remember that all this discussion was based on the assumption that $\gamma^\prime$ coupling to the SM charged particles vanish, making $\gamma^\prime$ (meta)stable.
If $\gamma^\prime$ coupling to the SM charged particles does not vanish, $\gamma^\prime$ will be unstable. We can then introduce another scalar $\phi^\prime$ with a coupling of form $\lambda_{C\phi^\prime}|C|^2(\phi^\prime)^2$. By imposing a $Z_2$ symmetry, $\phi^\prime$ becomes stable. By adjusting $\lambda_{C\phi}$, the $C\bar{C}\to \phi^\prime\phi^\prime$ annihilation mode
can be dominant.  The above argument can be then repeated by just replacing $\gamma^\prime$ with $\phi^\prime$.

Let us now enumerate the various bounds on $q$.
\begin{enumerate}
\item  {\it SLAC bound}: The
SLAC millicharged experiment has searched for hypothetical millicharged particles that could be produced at the SLAC positron target by looking for the ionization and excitation effects during the passage of the potentially produced millicharged particles in matter. The null results from the searches   set an upper limit of $4.1 \times 10^{-5}e - 5.8 \times 10^{-4} e $ on the electric charge  for millicharged particles of mass $1 ~\rm{MeV}  - 100 ~\rm{ Me}V$ \cite{Prinz:1998ua}.

\item  {\it Supernova bound} :  The main  process for milli-charged particle production in a supernova is plasmon decay
to these particles. Considering energy loss rate of Supernova $1987$A, an upper bound of $10^{-9}$ on charge is obtained
\cite{Davidson:2000hf}.

\item {\it Bounds from BBN}: As we discussed earlier, a  mixing between $\gamma^\prime$ and $\gamma$ can lead to the production of the $C$ and $\gamma^\prime$ particles in the early universe which can in principle affect BBN. In the context of the mirror model where each SM particle has a dark counterpart with the same mass, Ref \cite{Berezhiani:2012ru} finds an upper bound of
    $3 \times 10^{-10}$ on kinetic mixing (corresponding to an upper bound of $e\times 3\times 10^{-10}=10^{-10}$ on $q$). This bound does not apply in our case because in our model, the dark sector is not an exact replica of SM sector. As discussed before, in our model $C\bar{C}$ can be produced via Drell-Yann process $e^-e^+ \to \gamma^* \to C\bar{C}$ and $C\bar{C}$ in turn annihilate to a $\gamma^\prime$ pair. Notice that since $
    \gamma^\prime$ particles do not couple to the SM charged particles at tree level, they cannot be produced directly. (See appendix A of \cite{Feldman:2007wj}.)
    As discussed before taking $m_{\gamma^\prime}\ll$~MeV, the produced $\gamma^\prime$ will count as extra relativistic degrees of freedom; {\it i.e.,} a new contribution to $N_{eff}$. Taking $\Delta N_{eff}<0.7$ \cite{Kamada:2015era}, we find the limit shown in Fig 1 and marked with BBN.
\end{enumerate}

Remember that in our model the DM particles, $X$, have a  mass of $O(10~{\rm MeV})$. As mentioned before in the introduction, one plausible possibility  for the DM production in the early universe is the SLIM scenario \cite{Boehm:2006mi,Farzan:2009ji} in which $X$ particles are produced thermally with $S$-wave annihilation to neutrino pairs and anti-neutrino pairs: $\langle \sigma (X+X \to \nu \nu , \bar{\nu}\bar{\nu}) v \rangle|_{tot}=1$ pb. The bounds from CMB on $N_{eff}$ then implies  $m_X>5$ MeV \cite{Wilkinson:2016gsy}.

There are also bounds on the initial energy of $e^+$ injected in the galaxy, $E_{inj}$. If positrons produced in dark matter annihilation in galactic center are mildly relativistic, high energy gamma rays will be produced. Comparing gamma-ray spectrum due to inflight annihilation to the observed diffuse Galactic gamma ray data, an upper bound of $ 3 ~\rm{MeV}$  is found on $E_{inj}$ \cite{Beacom:2005qv}.  Notice however that this analysis assumes negligible ionization in galaxy. Taking into account the radiation components related to in-flight annihilation and internal
Bremsstrahlung with more conservative assumptions about ionization fraction,  Ref  \cite{Sizun:2006uh} sets an upper bound of $7.5 ~\rm{MeV}$ on $E_{inj}$. Another upper bound of $10~\rm{MeV}$ is found on $E_{inj}$  by considering Voyager spacecraft electron positron data  \cite{Boudaud:2016mos}. \footnote{The Voyager spacecraft  has crossed the heliopause in summer 2012 beyond which it can probe possible $e^{\pm}$ from sub-GeV DM without hinderance from solar magnetic field  \cite{Boudaud:2016mos}.}
Remember that in our model, $C\bar{C} \to \phi \phi$ is followed by $\phi \to e^-e^+$. Also remember that the energy of majority of $C$ particles is $m_C(1+v_f^2/2)\simeq m_C$. $E_{inj}$ is therefore expected to be $\sim m_C/2$.  The upper bound on $E_{inj}$ therefore corresponds  to an upper bound on $m_C/2$ which is shown in Fig $1$-b.

The $C$ and $\bar{C}$ particles can be easily trapped in the magnetic field of pulsars. However, the $e^+$ flux from $C\bar{C}$ annihilation in pulsars is negligible in comparison to diffuse flux. Let us study the effect of magnetic field of the sun on $C$ particles. The magnetic field at the surface of the sun is about $1$ Gauss \cite{Nghiem:2006wp}. Larmour radius of the trajectory of  $C$ particles  at the solar surface is then given by:
\be
r_L = 1.5 \times 10^{9} ~\rm{km} \left( \frac{3 \times 10^{-11}}{q} \right) \left( \frac{m_X}{5 ~\rm{MeV}} \right) \left( \frac{1 ~\rm{Gauss}}{B} \right).
\ee
Comparing it with the solar radius ($r_s = 7 \times 10^5$ km), we find that $r_s \ll r_L $. Thus, the effect of magnetic field of the sun on $C$ particles is negligible. The effect of  the magnetic field of Earth on the trajectory of $C$ particles is then by far negligible $(r_s \ll r_L )$.

\section{Direct detection}

  As is well-known, the recoil energy in the scattering of non-relativistic dark matter of mass of few MeV off nuclei is too small to be detectable by current direct dark matter search experiments.
   However, as we discussed in the previous section, $C$ particles are expected to have velocities of $v\sim 1/3$ in our vicinity (see Eq. (\ref{vf})).
    The $C$ particles, being relativistic can lead to average recoil energy  \be \langle \Delta E\rangle \sim 0.035~{\rm keV} \frac{1}{1-v^2}  \left( \frac{  v}{1/3} \right) ^2 \left( \frac{m_C }{3~\rm{MeV}} \right)^2 \left( \frac{28~\rm{GeV}}{M_N}\right)\left(\frac{M_N}{M_N+\gamma m_C} \right)^2\label{averageER}\ee at scattering off nuclei with mass $M_N$ which is below the energy threshold of current DM search experiments such as the LUX experiment \cite{Akerib:2015cja} with $E_{th}=3$ keV, DAMA \cite{Bernabei:2016wvk} with $E_{th}=2$ keV and CRESST II \cite{Schieck:2016nrp} with $E_{th}=300$ eV. However, the upcoming direct dark matter search experiments such as   SuperCDMS with $E_{th}=56$ eV \cite{Agnese:2017jvy}, CRESST III  with $E_{th} = 10-20$ eV \cite{Strauss:2017woq} and Edelweiss III with $E_{th}< 100$  eV \cite{Kozlov:2017laq} can probe such small recoil energies opening up the hope to probe this scenario.
     The target usually used in low energy threshold experiments is Silicon with mass of $28$ GeV. The differential event rate per unit of target mass
resulting from the pico-charged particle interactions can be written as
    \be \frac{dN_{events}}{dE_r}=\frac{1}{M_N} \frac{\rho _{X}}{  m_X} ( 2fv  ) \frac{d \sigma   }{dE_r } \ee
    where
    $$     \frac{d  \sigma }{ dE_{r}}= \mid F \mid^2 \frac{Z^2}{4 \pi M_N ^2} \times \frac{ e^2 q^2 m_C^2}{ E_r ^2} \times \frac{1}{E_r ^{max}}
$$
in which $E_r^{max}=2\langle \Delta E\rangle \sim 70 $ eV is maximum recoil energy.
    In above formula,  $\mid F \mid$ is the nuclear form factor and is close to one when momentum transfer is much smaller than the nucleus target mass.
    The event rate can be written as
        \be N_{events}= 0.001  \left( \frac{q}{10^{-11}} \right)^2  \left( \frac{f}{10^{-4}} \right)  \left( \frac{m_C}{3~\rm{MeV }} \right)^2 \frac{ \rm{70~{\rm eV}}}{E_r ^{max }}  \left(\frac{56 ~{\rm eV}}{E_{th}}-\frac{70 ~{\rm eV}}{ E_r^{max}}\right) \left( \frac{10 ~\rm{ MeV }}{m_X} \right) \left( \frac{28 ~\rm{GeV }}{M_N} \right)^3 (\rm{kg.year})^{-1} .
         \ee
    In the above formula, we have set $E_{th}=56$ eV which is the value reported for SuperCDMS \cite{Agnese:2017jvy}.
The current bound from CRESST II  constrains $N_{events}<0.1 ~(\rm{kg.day})^{-1}$ at $1\sigma$ C.L. \cite{Brown:2011dp}. A slightly weaker  bound on the event rate is provided by  the Edelweiss II experiment \cite{Armengaud:2009hc}: $N_{events}<0.16 ~(\rm{kg.day})^{-1}$ at $1\sigma$. The corresponding bounds on the charge are shown in Fig. $1$-b for $m_X=10$~MeV and $f=10^{-4}$. Pico-charged $C$ particles with electrical charge of $10^{-11}$ can be detected with  future experiments if the bound on $N_{events}$ is improved by a factor of $100$.
The next phase of  Edelweiss III experiment will start  data taking in $2018$ with an exposure of $1200$ kg.days with energy threshold of less than $100$ eV \cite{Kozlov:2017laq}. The curve in figure $1$-b  marked with Edelweiss III shows the forecast for the sensitivity of the Edelweiss III experiment with $E_{th} = 60$ eV.
The SuperCDMS experiment   will start data taking in $2020$ and is expected to achieve an exposure of $1800$ kg.days \cite{Agnese:2016cpb}. As can be seen from the plot for this exposure and $E_{th} = 56$ eV this experiment will be able to probe a significant  part of the parameter space of our model.
 Phase II-b CRESST III experiment will start  data taking in January $2018$ for expected time scale of one year with exposure of  $1000$ kg.days \cite{Angloher:2015eza}. Phase III of CRESST III experiment will  take data from January $2019$ to December $2020$ reaching very low energy threshold of $10-20$ eV \cite{Angloher:2015eza}. As is shown in  figure $1$-b, CRESST III experiment with $E_{th} =10$ eV  and with an exposure of $1000$ kg.days (which can be achieved by $2020$) can probe the entire parameter space of our interest.

Notice that in our model $d \sigma/dE_r \propto 1/E_r ^2 $ which indicates that when recoil energy decreases, $d \sigma/dE_r$
increases as $E_r ^{-2}$. This energy dependence is very unique and is not expected in dark matter models that interact with matter via exchange of a heavy virtual state. Even in light anapole or magnetic
dipole dark matter models, where interaction is through exchange of the photon, we do not expect such a dependence on $E_r$ because in these models the vertex is also proportional to energy-momentum exchange \cite{DelNobile:2017fzy}. Thus, if statistics allows, by reconstructing the dependence of $d\sigma/dE_r$ on $E_r$, this scenario can be tested.

Before summarizing, let us  study the possibility of studying the scattering of the $C$ particles off the electrons in detector. Replacing $M_N$ with $m_e$ in Eq. (\ref{averageER}), we arrive at  the formula for average recoil energy scattering off electrons in a target which gives  $\langle \Delta E\rangle|_{electron}=54$ keV. The cross section is given by
\be \int_{E_{th}}\frac{d \sigma_e}{dE_r} dE_r= \frac{\alpha q^2}{m_e^2}\frac{1}{v^4}\left(\frac{m_e v^2}{E_{recoil}}-\frac{1}{2} \right)=10^{-43}~{\rm cm}^2\left(\frac{0.3}{v}\right)^2\left(\frac{q}{10^{-11}}\right)^4. \ee
This is well below the current sensitivity limit of XENON$100$ \cite{Essig:2017kqs} and leads to a weak constraint on electrical charge ($q \stackrel{<}{\sim} 10^{-7} $ for $m_c \sim 3$  MeV).
It may be probed by a larger Borexino-type solar neutrino experiment with an energy threshold  below $\sim 50$~keV. Studying the potential effect is beyond the scope of current paper and will be studied elsewhere.

Let us summarize the main results of this section.
 Taking into account the background for $N_{events}$, the bounds from these experiments are shown in Fig. 1-b.
 As seen from this figure, the $C$ particles with mass larger than $10$ MeV coming from decay of dark matter with mass of $40$ MeV, are already excluded by CRESST II experiment. Future experiments such as Edelweiss III, CRESST III and SuperCDMS will be able to probe a significant part of the parameter space. For $C$ particles coming from decay of dark matter with mass of $5$ MeV (lower bound on dark matter mass from CMB bound), CRESST III experiment with energy threshold of $10$ eV, will be able to probe the entire parameter space of relevance to this model.
\begin{figure}[t]
\includegraphics[scale=0.35]{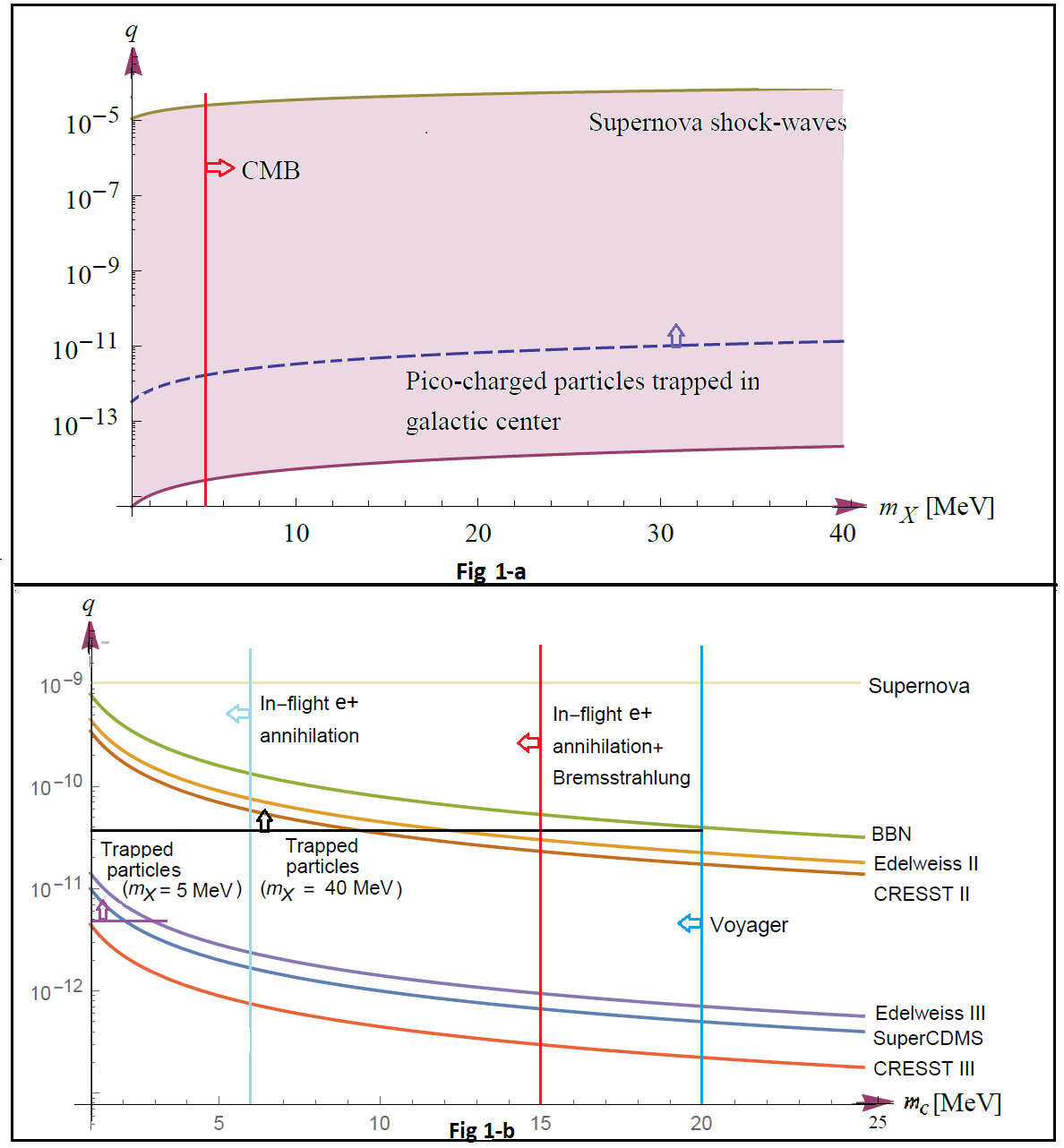}
\caption{\textbf{Fig a (upper panel):} Electric charge of $C$ particles versus dark matter mass, $m_X$, which is equal to twice the energy of $C$ particles at production in the galaxy: $E_C=m_X/2$. Blue dashed curve shows lower limit above which the $C$ particles with energy $m_X/2$ stay in magnetic field of 10 $\mu$G with Larmour radius below 100~pc. Red vertical line indicates CMB lower bound on dark matter mass \cite{Wilkinson:2016gsy}. Shaded area shows the range of $q$ for which energy pump by supernova shock-waves in the galaxy is efficient.
\textbf{Fig b (lower panel):} Bounds on the charge of $C$ particle versus its mass. Line marked with supernova comes from supernova cooling \cite{Davidson:2000hf}. Curve marked with BBN comes from imposing $\Delta N_{eff}<0.7$ at the BBN era.  Current bounds from Edelweiss II \cite{Kozlov:2017laq} and CRESST II \cite{Schieck:2016nrp} direct dark matter search experiments are shown along with the forecast for future experiments: Edelweiss III with   $E_{th}= 60$ eV,  SuperCDMS with $E_{th}=56$ eV and CRESST III with $E_{th}=10$ eV. Curves are drawn  under conservative assumption $f = \Gamma_X t^0= 10^{-4}$ and $m_X=10$~MeV. The velocity of the $C$ particles is taken equal to $1/3$.  The horizontal lines at $4\times 10^{-11}$ and $5\times 10^{-12}$ show lower bound on $q$ above which $C$ particles with energies respectively equal to $m_X/2=20$ MeV and $m_X/2=2.5$ MeV do not leave the vicinity of the Earth; that is, they have Larmour radius below 0.3 kpc (the thickness of the galactic disk) for $B=1~{\mu G}$ (typical magnetic field at 8 kpc from galactic center). Vertical lines show upper bounds on $m_C$ (twice the energy of $e^+$ at injection) from voyager \cite{Boudaud:2016mos} and different analysis of inflight annihilation  \cite{Beacom:2005qv,Sizun:2006uh}.}
\end{figure}


\section{Summary and concluding remarks}
We have proposed a novel scenario for dark matter within which the chances for finding an indirect dark matter signal from dark matter in our galaxy enhances despite the new stringent bounds from delayed recombination CMB bound and null results from dwarf galaxies. In this scenario, dark matter is composed of scalar metastable particles (denoted by $X$) which decays to a pair of $C$ and $\bar{C}$ particles with a lifetime of $\sim 1000$ times the age of the universe. The $C$ and $\bar{C}$ particles are charged under a new $U(1)_X$ gauge symmetry whose gauge boson called dark photon mixes with the hypercharge gauge boson creating a tiny electric charge for the $C$ particles. The $C$ and $\bar{C}$ particles, produced by the decay of particles in galaxy, are trapped in  a distance from galaxy center close to that of their production point by the galactic magnetic field. Their eventual annihilation  leads to indirect dark matter signal.

At production the $C$ particles are ultrarelativistic ($v\simeq 1$) but as we showed, their scattering off the background relic dark photons provides an energy loss mechanism which reduces their velocity in the galaxy down to $v\sim 1/3-1/2$. This natural energy loss mechanism protects the $C$ and $\bar{C}$ content of galaxy against energy pump by supernova shock waves \cite{Chuzhoy:2008zy}.

The upper bound on the ratio of relativistic $C$ and $\bar{C}$ particles to $X$ particles ({\it i.e.,} $f=\Gamma_X t^0$) is $10^{-3}$. In this model, the density of $C$ and $\bar{C}$ particles at recombination would be too small to lead to a significant delay in recombination simply because at $ \Gamma_X t|_{recombination} \ll 1$. Moreover, the magnetic fields at dwarf galaxies are too weak to keep
$C$ and $\bar{C}$ from escape so we do not expect DM signals from dwarf galaxies.

The main focus of the present paper was on a mechanism to revive the DM interpretation of the 511 keV line observed from the galaxy center. In order to account for the intensity of the line, the
$C \bar{C}$ annihilation rate should be given by $50~{\rm pb} (0.0001/f)^2(m_X/5~{\rm MeV})^2$. The energies of $e^+$ and $e^-$ at injection are expected to be around $m_C/2$ so the upper bounds
on $E_{inj}$ from various observation can be translated into bounds on $m_C<{\rm few} ~ 10$ MeV (independent of the value of $m_X$) as shown in Fig. $1$-b.
The $C$ and $\bar{C}$ particles can be trapped in the magnetic field of pulsars; however, the 511 keV flux at Earth from annihilation of the $C\bar{C}$ particles in the pulsars (point sources) is expected to be negligible compared to the diffuse flux from galaxy center.

 The $C $ and $\bar{C}$ particles, having electric charge, can scatter off protons in the direct dark matter search experiments. Since these particles are relativistic ($v\sim 1/3-1/2$), the recoil energy can be relatively large and discernible at upcoming direct dark matter search experiments with $E_{th}<100$ eV such as Edelweiss III,  superCDMS and CRESST III.  As shown in Fig $1$-b, the entire parameter space of relevance for the 511 keV solution can be tested by these experiments by circa 2022.
If the mass of $C$ particles is relatively large $\sim 15-20$ MeV, the $e^+$ particles from its annihilation may be detected by the Voyager spacecraft which is now exploring the interstellar medium \cite{Boudaud:2016mos}. Another characteristic prediction of the scenario is the absence of a signal from dwarf  galaxies or from parts of DM halo beyond the disk where magnetic field is weak.
The magnetic fields at dwarf galaxies are still poorly known. With better studying the correlation of the 511 keV signal with  magnetic fields, it will be possible to test this model. Measurement of the  magnetic field in dwarf galaxies is possible  by Faraday rotation and polarization measurements \cite{Beck:2008ty}.
Future high resolution and  high sensitivity observations at high radio frequencies with
the Jansky Very Large Array (VLA) and the planned Square Kilometer Array  will make it possible to study the magnetic field structure of dwarf galaxies more precisely \cite{Beck:2013bxa}.

\subsection*{Acknowledgments}
Authors would like to thank Z.~Liu for useful comments. The authors are also grateful to Neal Weiner and Ben Kilminster for encouraging remarks. They are also thankful to the anonymous referee for her/his very useful remarks.
This project has received funding from the European Union\'~\!s Horizon 2020 research and innovation programme under the Marie Sklodowska-Curie grant agreement No 674896 and No 690575.
YF is also grateful ICTP associate office and Iran National Science Foundation (INSF) for partial financial support under contract 94/saad/43287.



\end{document}